\documentclass[twocolumn,twocolappendix]{aastex7}
\usepackage{CJK} \usepackage{soul} \usepackage{amsmath}
\allowdisplaybreaks[4]

\newcommand{\G}{\mathcal{G}}

\newcommand{\kms}{\ensuremath{\rm km\,s^{-1}}}

\newcommand{\appropto}{\mathrel{\vcenter{\offinterlineskip\halign{\hfil$##$\cr\propto\cr\noalign{\kern2pt}\sim\cr\noalign{\kern-2pt}}}}}
\newcommand{\bjdtdb}{\ensuremath{\rm {BJD_{TDB}}}}
\newcommand{\feh}{\ensuremath{\left[{\rm Fe}/{\rm H}\right]}}
\newcommand{\teff}{\ensuremath{T_{\rm eff}}}
\newcommand{\logg}{\ensuremath{\,{\rm log}\,{g}}}
\newcommand{\ecosw}{\ensuremath{\sqrt{e}\cos{\omega_*}}}

\newcommand{\esinw}{\ensuremath{\sqrt{e}\sin{\omega_*}}}

\newcommand{\mstar}{\ensuremath{\,M_{\rm *}}}
\newcommand{\rstar}{\ensuremath{\,R_{\rm *}}}

\newcommand{\mratio}{\ensuremath{\,M_{\rm P}/M_{*}}}
\newcommand{\mplanet}{\ensuremath{\,M_{\rm P}}}
\newcommand{\rp}{\ensuremath{\,R_{\rm P}}}

\newcommand{\vsini}{\ensuremath{v\sin{i_*}}}
\newcommand{\kepler}{{\it Kepler}}
\newcommand{\ktwo}{{\it K2}}
\newcommand{\tess}{{\it TESS}\,}
\newcommand{\gaia}{{\it Gaia}}

\newcommand{\corot}{{\it CoRoT}}

\graphicspath{{./}{Figures/}}

\newcommand{\dotarcsec}{\rlap{.}''}

\newcommand{\nsysfitted}{256}
\newcommand{\nsysfittedwrm}{234}
\newcommand{\ndtornorm}{22}
\newcommand{\nsysusedforanalysis}{146}
\newcommand{\nss}{30}
\newcommand{\njup}{84}
\newcommand{\nmassive}{32}
\newcommand{\nbadplmass}{6}
\newcommand{\nbin}{90}

\newcommand{\nsysusedforanalysisworm}{10}
\newcommand{\nssworm}{4}
\newcommand{\njworm}{2}
\newcommand{\nmjworm}{4}

\usepackage{url}
\usepackage{amsmath}
\usepackage{float}
\usepackage{rotating}
\usepackage{gensymb}
\usepackage{hyperref}
\usepackage{pgffor} \usepackage{soul}
\usepackage{changepage}
\usepackage{xfp} 

\begin{document}
\begin{CJK*}{UTF8}{gbsn}

\title{A Homogeneous Catalog of Rossiter-McLaughlin Systems: \\Distinct $e-\lambda$ Trends in Three Gas-Giant Mass Regimes}

\shortauthors{Wang et al.}

\author[0000-0002-0376-6365]{Xian-Yu Wang} \altaffiliation{Sullivan Prize Postdoctoral Fellow}
\email{xwa5@iu.edu}
\affiliation{Department of Astronomy, Indiana University, Bloomington, IN 47405, USA}

\author[0000-0002-7846-6981]{Songhu Wang} \email{sw121@iu.edu}
\affiliation{Department of Astronomy, Indiana University, Bloomington, IN 47405, USA}

\author[0000-0002-7094-7908]{Konstantin Batygin} \email{kbatygin@caltech.edu}
\affiliation{Division of Geological and Planetary Sciences, California Institute of Technology, Pasadena, CA 91125, USA}

\correspondingauthor{Xian-Yu Wang}
\email{xwa5@iu.edu}
\correspondingauthor{Songhu Wang}
\email{sw121@iu.edu}

\begin{abstract}
Stellar obliquity ($\lambda$) and orbital eccentricity ($e$) trace the dynamical histories of close-in giant planets, but the current observational picture is assembled from heterogeneous analyses that have obscured population-level trends. In this work, we homogeneously refit systems with Rossiter-McLaughlin (RM) measurements by performing a joint global fit to spectral energy distributions, transit light curves, mid-transit times, out-of-transit and in-transit radial velocities, yielding self-consistent posterior distributions for the physical and orbital parameters of both stars and planets across {\nsysfitted} systems\footnotemark. Restricting to {\nsysusedforanalysis} single-star systems with reliable planet-mass measurements, we uncover pronounced structure in the $e-\lambda$ plane that depends on planet mass: (i) sub-Saturns ($M_{\rm p} \leq \sim0.3M_{\rm J}$) can be both eccentric and misaligned; (ii) Jupiters ($\sim0.3M_{\rm J}<M_{\rm p} \leq \sim3 M_{\rm J}$) are misaligned only on circular orbits; and (iii) {super Jupiters} and brown dwarfs ($M_{\rm p}>\sim3M_{\rm J}$) are aligned across the full eccentricity range. A two-dimensional Kolmogorov-Smirnov test shows that the joint $(e,\lambda)$ distributions differ significantly among these three mass regimes. These trends demonstrate that $\lambda$ depends jointly on eccentricity and planet mass, implying that obliquity alone is not a unique tracer of evolutionary history and underscoring the need for a unified framework for the origins of spin-orbit misalignment.
\end{abstract}

\keywords{exoplanet systems (484), exoplanet dynamics (490), exoplanets (498), planetary alignment (1243), planetary theory (1258), star-planet interactions (2177)} 

\footnotetext{For completeness, our sample also includes \ndtornorm\, systems whose \\ projected stellar obliquities were inferred solely from line-profile \\ techniques, for which the relevant data products are often not \\ uniformly accessible, or from analyses without publicly available \\ RM data. As a result, our global analysis for these systems is \\ limited to deriving self-consistent system parameters, while the \\ spin-orbit angles are adopted from the published literature. The \\full catalog will be publicly available at \url{www.stellarobliquity.com}.}

\section{Introduction} 

The first planets to be discovered beyond the Solar System were the close-in giants whose short periods made them unmissable to early radial-velocity surveys. Their existence immediately raised a profound question that still has no fully satisfactory answer: how did these massive worlds end up skimming the surfaces of their stars? Because these planets occupy a relatively narrow band of orbital separations, their semimajor axes provide little leverage on the physics of their origins (except for the tidally shaped pile-up of hot Jupiters; see, e.g. \citealt{Cumming1999, Udry2003, Gaudi2005, Wu2018}). Instead, the dynamical clues are carried by the remaining degrees of freedom. Accordingly, eccentricity and inclination distributions of close-in giants are often treated as fossil records of the processes that shaped them, potentially encoding traces of disk-driven migration, high-eccentricity past, and tidal evolution.

Historically, distinct formation and migration pathways for close-in giant planets were associated with different predictions for their dynamical architectures. \textit{In-situ} assembly \citep{Batgyin2016, Boley2016} and smooth disk migration \citep{Goldreich1980, Lin1996} were expected to maintain low eccentricities and small inclinations, while high-eccentricity channels such as planet-planet scattering \citep{Rasio1996, Weidenschilling1996}, secular interactions \citep{Wu2011, Petrovich20152015CHEM, Teyssandier2019}, and Kozai-Lidov cycling \citep{Holman1997, Wu2003, Fabrycky2007, Naoz2016, Vick2019} were envisioned as natural pathways to generate large values of both $e$ and $i$ prior to tidal damping. This contrast inspired the long-standing expectation that \textit{individual} diagnostics, particularly high eccentricity \citep[e.g.,][]{Naef2001HD80606, Brady2018, Dong2021, Jackson2023, Gupta2024} or large mutual inclinations \citep{Mills2017, Masuda2017, Almenara2022, Nabbie2025}, could be used to distinguish between formation channels, under the assumption that these dynamical signatures covary at least in a statistical sense. Observers have often relied on measurable proxies for these quantities, such as orbital isolation \citep{Steffen2012, Huang2015, Becker2015, Canas2019, Wu2023HJsNotAlone} as an indicator of a high-$e$ past, or the 3-D stellar obliquity $\psi$ \citep[e.g.,][]{Winn2010, Schlaufman2010, Albrecht2012, Albrecht2022, Knudstrup2024} as a surrogate for mutual inclination.

A growing body of theory shows that this one-to-one mapping between mechanisms and observables is rarely preserved. Eccentricities can be excited without significantly altering spin-orbit alignment through co-planar secular resonances \citep{Li2014EccentricityGrowth, Petrovich20152015CHEM}. Conversely, stellar obliquities can be tilted while leaving eccentricities essentially unchanged. Proposed sources of such misalignments include primordial star-disk tilting driven by chaotic accretion \citep{Bate2010, Thies2011, Fielding2015, Bate2018}, magnetic warping \citep{Foucart2011, Lai2011, Romanova2013, Romanova2021}, or internal gravity wave-driven stellar tumbling \citep{Rogers2012}, as well as disk torquing by stellar companions \citep{Batygin2012,Matsakos2017, Su2025}. Tidal evolution adds further complexity because the characteristic damping timescales for $e$ and $\psi$ differ in both magnitude and functional dependence \citep{Lai2012, Winn2010, Albrecht2012, Li2016, zanazzi2024damping}. As a result, tides may efficiently erase eccentricity while leaving the stellar spin misaligned, or conversely realign the star while preserving a remnant of the planet's dynamical past.

As a means toward unraveling this complexity, previous studies have primarily linked stellar obliquity to \emph{stellar} properties, with the most notable trends appearing across the $T_{\mathrm{eff}}$ boundary \citep{Schlaufman2010, Winn2010, Wang2025kb}. While not immediately expected, there is no a priori reason that \emph{planetary} properties cannot be equally important. Accordingly, we treat these as hitherto under-explored degrees of freedom and search for empirical structure in the $e$-$\psi$ plane, recognizing that $e$ and $\psi$ need not rise and fall together but can retain specific, correlated imprints of distinct dynamical pathways.

\begin{figure*}[t]
    \centering
    \includegraphics[width=1\linewidth, trim=0 0 200 0, clip]{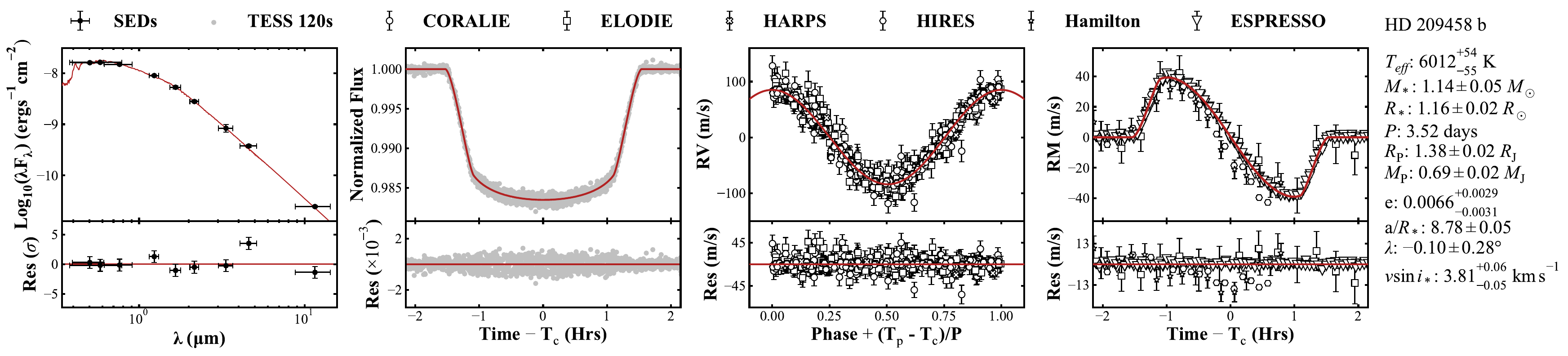}
    \caption{An example plot for the global modeling of HD 209458, the first exoplanetary system with RM measurement \citep{Queloz2000}. From left to right, each panel represents the fits to the spectral energy distribution (SED), \tess\, transits, radial velocities, and Rossiter-McLaughlin effects. Figures for all systems can be found \href{https://github.com/wangxianyu7/Data_and_code/tree/main/e-\%CE\%BB_Trends}{here}.}
    \label{fig:example_modelling}
\end{figure*}

A range of statistical techniques can be used to infer eccentricity from the photo-eccentric effect \citep{Dawson2012,Kipping2013eccq1q2}, estimate mutual inclinations from multiplicity statistics \citep{Lissauer2011, Tremaine2012}, transit-duration distributions \citep{Fabrycky2014}, and transit-timing and transit-duration variations (TTVs/TDVs, \citealt{Zhu2018, Millholland2021}). Simultaneously, line-of-sight stellar obliquity ($i_\star$) can be constrained from $v\sin i_\star$ statistics \citep{Schlaufman2010, Winn2017, Louden2021, Louden2024, Morgan2024, Biddle2025} or from the amplitude distribution of spot modulation in light curves (e.g., \citealt{Mazeh2015}). Nevertheless, the statistical power of these ensemble methods rest chiefly on large samples of detected small planets \citep{Xie2016,vaneylen2019ecc,Zhu2018,Millholland2021,Gilbert2025}. Consequently, they have historically offered limited population-level constraints on the orbital architecture of giant planets because of their low occurrence rate, although recent studies have begun to assemble larger samples of giant planets \citep{Dong2021WJ, Jackson2023,Fairnington2026}.

Decisive constraints for giant planets, therefore principally arise from direct, system-by-system measurements, i.e., out-of-transit radial velocities for eccentricity and in-transit spectroscopy (the Rossiter-McLaughlin effect, RM effect; \citealt{Rossiter1924,McLaughlin1924}) for the sky-projected spin-orbit angle which serves as an established observational proxy for the true obliquity $\psi$ \citep{Albrecht2012,Dong2023StellarObliquity, Siegel2023,Knudstrup2024,Bourrier2023}. Even such direct measurements, however, mix results from heterogeneous analyses with inconsistently modeled uncertainties, obscuring population-level inferences. 

On the eccentricity side, reported eccentricities often suffer from the Lucy-Sweeney bias  (since the zero phase lies exactly at $e=0$, observational uncertainties tend to bias the best-fit value upward; \citealt{Lucy1971}). This bias is not easily corrected a posteriori,  because its magnitude depends on the radial-velocity signal-to-noise ratio, the time span of the observations \citep{shen_eccentricity_2008}, the phase coverage \citep{zakamska_observational_2011}, and additional factors such as long-term trends, correlated noise, and non-uniformly handled jitter \citep{hara_bias_2019}. Moreover, some literature works have explicitly assumed circular orbits ($e = 0$), which not only removes the observational uncertainty on eccentricity but also risks misrepresenting systems whose true orbits may deviate from circularity, thereby biasing population-level analyses (see \citealt{Morgan2025} for a recent uniform RV-based reassessment of warm-Jupiter eccentricities). 

On the RM side, different groups adopt different forward models \citep{Ohta2005,Hirano2010,Hirano2011,Covino2013,Boue2013} and modeling strategies, introducing systematic biases in both $\lambda$ and \vsini\ \citep{Hirano2011,Brown2017,Triaud2018,Brady2025}. Similar to $e$ and $\omega$, \vsini\ and $\lambda$ are highly correlated. This degeneracy is especially severe when \vsini\ is small or poorly constrained, and when the transit geometry (e.g., impact parameter) is poorly constrained, as is common in legacy ground-based discoveries lacking space-based photometry, leaving $\lambda$ ill-constrained \citep{Triaud2018}. Furthermore, RM-only modeling (i.e., fitting the RM signal in isolation rather than in a joint global model including transits and RVs) and the early reliance on Monte Carlo or bootstrapping methods (prior to the widespread adoption of MCMC and nested sampling) can fail to propagate uncertainties and covariances correctly. Collectively, these issues have fueled debate over whether any $e$ -- $\lambda$ relation is real or merely an artifact of biases and systematics \citep{Rice2022,Knudstrup2024,Gan2026}, despite intriguing results suggesting that eccentric Jupiters tend to be aligned \citep{Espinoza2023} (see Figure~\ref{fig:elambdarelation} and Section~\ref{section:e_lambda} for our mass-stratified view).

To address these issues, we perform a homogeneous global analysis of \nsysfitted\, systems with RM measurements, jointly modeling spectral energy distributions, transit light curves and mid-transit times, and both out-of-transit radial velocities and in-transit RMs, with a consistent, modern treatment of $e$ and $\lambda$ (see Section~\ref{section:global_fit}). This homogeneous refit yields self-consistent posteriors for $e$, $\lambda$, and the full set of orbital and physical parameters and reveals pronounced structure in the $e-\lambda$ plane for single-star hosts with robust planet-mass measurements: Sub-Saturns can be both eccentric and misaligned; Jupiters are misaligned only on circular orbits; and {super} Jupiters and brown dwarfs are aligned across the full eccentricity range (See Section~\ref{section:e_lambda}).

The paper is organized as follows. Section~\ref{sec:Modeling Sample Construction and Data Collection} describes the sample construction and data collection. Section~\ref{section:global_fit} presents the global modeling. Section~\ref{section:e_lambda} presents the key results, and Section~\ref{sec:discussion} offers the discussion.

\section{Modeling Sample Construction and Data Collection}
\label{sec:Modeling Sample Construction and Data Collection}
\subsection{Modeling Sample Construction}

Stellar obliquities can be constrained through several complementary techniques, including transit-based, spatially resolved methods such as the RM effect and its variants, starspot-crossing and gravity-darkened transits, as well as inclination-based methods such as asteroseismology or by inferring $i_\star$ from $(R_\star, P_{\rm rot}, v\sin i_\star)$. 

In practice, 339 out of 468 obliquity measurements\footnote{TEPCat, \citealt{Southworth2011,tepcat2026}, \url{https://www.astro.keele.ac.uk/jkt/tepcat/obliquity.html}
. Accessed on Jan 1, 2026.} for individual planet-hosting stars are RM-based. We therefore perform homogeneous global modeling for each system to uniformly derive the physical and orbital parameters of both stars and planets. For \nsysfittedwrm\ systems with classical RM time-series RVs, we measure the projected obliquity by fitting the in-transit RM signal following \citet{Hirano2011}.

For \ndtornorm\, systems whose RM data are unavailable, or for which projected obliquities were derived from line-profile techniques, including Doppler shadow/tomography \citep{Albrecht2007, Collier2010, Zhou2016, Johnson2017}, reloaded RM \citep[RRM;][]{Cegla2016ReloadedRM}, and RM Revolutions \citep[RMR;][]{Bourrier2021RMrevolutions}, we use our global fits to provide self-consistent system parameters and adopt the published spin--orbit angles in the subsequent statistical analysis.

For some young or active stars, modeling the RM signal requires a joint treatment of stellar activity, commonly implemented with Gaussian processes. As this lies beyond the scope of this study, we exclude these systems to preserve a uniform analysis framework. We also remove systems with transit timing variation amplitudes $>$ 5 minutes, which require photodynamical modeling \citep{Doyle2011,Carter2012, Wang2018a} . The fitted \nsysfitted\, systems have been listed in Table~\ref{tab:listandparameters}.

\subsection{Data Collection}
\subsubsection{Transits and Mid-transit Times}

\noindent\uline{Ground-based transits}

We searched for ground-based photometry from the Strasbourg Astronomical Data Center (CDS)\footnote{\url{https://cds.unistra.fr/}}
 and the Exoplanet Follow-up Observing Program (ExoFOP)\footnote{\url{https://exofop.ipac.caltech.edu/tess/}}
. To efficiently compile photometry hosted at CDS, we first queried the NASA Exoplanet Archive using the TIC IDs of the systems in our sample to identify associated references. We then used the bibcodes of these references to check whether the corresponding publications provide photometric data archived at CDS. For ExoFOP, we downloaded all available photometry for the systems in our sample and removed data sets that have not yet been published. For the ExoFOP data, we adopted the default detrending configuration specified in the .plotcfg files provided on ExoFOP. The ground-based transits used in this work are listed in Table~\ref{tab:data}.

\noindent\uline{Space-based transits}

To provide precise constraints on transit profiles, we model light curves from space missions such as $\corot$ \citep{Auvergne2009}, $\kepler$ \citep{Borucki2010Sci}, $\ktwo$ \citep{Howell2014}, $\tess$ \citep{Ricker2015}, and \textit{Spitzer} \citep{Werner2004} for nearly all targets, using them to constrain transit-shape parameters ($R_{\rm p}/R_\star$, $a/R_\star$, $i$, $q_1$, and $q_2$\footnote{The transformed parameters ($q_1$, $q_2$) are related to the quadratic limb-darkening coefficients ($u_1$, $u_2$) via Equations 15 and 16 of \cite{Kipping2013eccq1q2}: $u_1 = 2\sqrt{q_1}\,q_2$ and $u_2 = \sqrt{q_1}(1 - 2q_2)$.}).

The $\corot$ data were retrieved from the IAS CoRoT Public Archive \footnote{\url{http://idoc-corot.ias.u-psud.fr/sitools/client-user/COROT_N2_PUBLIC_DATA/project-index.html}}. Light curves from $\kepler$, $\ktwo$, and $\tess$ were obtained using the \texttt{lightkurve} package \citep{Lightkurve2018}. For $\kepler$ data, 1-minute short-cadence light curves were prioritized; when unavailable, 30-minute long-cadence data were used. For $\ktwo$, we employed the 30-minute cadence light curves extracted using the K2 Self Flat Fielding pipeline \citep[K2FF,][]{Vanderburg2014K2SFF}. For \tess\, light curves, we followed the light-curve selection approach described in \citet{Ivshina2022}. When available, we used the Pre-search Data Conditioning Simple Aperture Photometry \citep[PDCSAP,][]{Smith2012PDCSAP, Stumpe2012, Stumpe2014} light curves, which were processed by the Science Processing Operation Center (SPOC) team \citep{Jenkins2016}. In cases where PDCSAP light curves were unavailable, light curves from the MIT Quick-Look Pipeline \citep[QLP,][]{Huang2020QLP} were considered. 

All space-based photometry was detrended using the \texttt{wotan} package \citep{Hippke2019}, applying the Cosine Filtering with Autocorrelation Minimization (CoFiAM) algorithm \citep{Kipping2013Cos} with a window size of three times the transit duration. The space-based transits used in this work are listed in Table~\ref{tab:data}.

\noindent\uline{Mid-transit times}

In global modeling, the transit data constrain both the transit-shape parameters and the orbital ephemeris ($P$ and $T_c$). Owing to their high photometric precision, space-based light curves are generally sufficient to constrain the transit-shape parameters even without including ground-based transits. However, the transit ephemeris benefits from a longer time baseline, and accurate ephemerides determine the expected transit mid-times of RM observations and thus improve constraints on $\lambda$. We therefore additionally fit literature ground-based transit mid-times compiled by the ExoClock Project\footnote{\url{https://osf.io/p298n/}}
 \citep{ExoClockI, ExoClockII, ExoClockIII}. 

All light curve and mid-transit time timestamps used in this work were converted to \bjdtdb, using the \texttt{barycorrpy} package\footnote{\url{https://github.com/shbhuk/barycorrpy}} \citep{barycorrpy, Eastman2010, Wright2014}. The source of the light curves for each target is detailed in Table~\ref{tab:data}. 

\subsubsection{Radial Velocities and RM measurements}

Radial velocity data for each planet were obtained from relevant literature sources. We thoroughly reviewed the references associated with each planet in the NASA Exoplanet Archive \citep{Akeson2013} to compile the available data. Moreover, for RM measurements, we reviewed the stellar obliquity references listed in the TEPCat to ensure that all RM measurements for each system were included. To ensure consistency with the transit data, all RM timestamps were converted to \bjdtdb\ using the \texttt{barycorrpy} package. The sources of RV data we adopted are also presented in Table~\ref{tab:data}.

\section{The global fit}
\label{section:global_fit}

To maintain consistency across stellar and planetary parameters and to propagate parameter uncertainties coherently, we used \texttt{allesfast}\footnote{\url{https://github.com/wangxianyu7/allesfast}}, which builds upon \texttt{allesfitter} \citep{allesfitter-paper, allesfitter-code} and incorporates selected features from \texttt{EXOFASTv2} \citep{Eastman2017, Eastman2019}, to simultaneously fit each system's spectral energy distributions (SED), transits, transit mid-times, and out-of-transit orbital radial-velocity data, as well as the in-transit Rossiter-McLaughlin measurements. In \texttt{allesfast}, we implemented an RM module following the formulation of \citealt{Hirano2011}, which accounts for stellar rotation velocity ($v_{eq}$ or $\vsini$), macroturbulence ($\zeta$), microturbulence ($\xi$), thermal broadening ($\beta$), pressure broadening ($\gamma$), and instrumental broadening ($\beta_{\rm IP}$).

\texttt{Allesfast} fits the observational data by maximising the total
log-likelihood. At each step, the objective function is the sum of the
following contributions:

\begin{align}
\ln\mathcal{L}_{\mathrm{total}} =\;
& \ln\mathcal{L}_{\mathrm{Transit}}
+ \ln\mathcal{L}_{\mathrm{RV}} \nonumber\\
& + \ln\mathcal{L}_{\mathrm{MIST}}
+ \ln\mathcal{L}_{\mathrm{SED}}
+ \ln\pi(\boldsymbol{\theta})
\end{align}
where the photometric and rv terms take the form
\begin{equation}
\ln\mathcal{L}_{\mathrm{data}} =
-\frac{1}{2}\sum_{i=1}^{N}
\left[\frac{(d_{i}-\mathcal{M}_{i})^{2}}{\sigma_{i}^{2}}
+ \ln\!\left(2\pi\,\sigma_{i}^{2}\right)\right],
\end{equation}
$\mathcal{M}_{i}$ is the physical model (including any baseline and
stellar variability), and $\sigma_{i}$ includes the fitted jitter.
The Rossiter--McLaughlin effect is evaluated as part of the RV model.
$\ln\mathcal{L}_{\mathrm{MIST}}$ constrains the stellar parameters
via MIST evolutionary tracks,
$\ln\mathcal{L}_{\mathrm{SED}}$ constrains the broadband magnitudes,
and $\ln\pi(\boldsymbol{\theta})$ collects Gaussian and uniform priors
on the fitted parameters as well as the linear-ephemeris prior on
transit mid-times.

\noindent\textit{\ul{MIST + SED fit}}

To derive the stellar parameters, including effective temperature (\teff), metallicity (\feh), surface gravity (\logg), mass (\mstar), and radius (\rstar), we performed a spectral energy distribution (SED) analysis by jointly fitting broadband photometry with the MESA Isochrones \& Stellar Tracks (MIST) evolutionary models \citep{Choi2016mist, Dotter2016mist}.

The broadband photometry for each star was compiled from various catalogs, including 2MASS $J$, $H$, and $K$ bands \citep{Cutri2003}, WISE $W1$, $W2$, and $W3$ bands \citep{Cutri2014AllWISE}, and Gaia $G$, $G_{\rm BP}$, and $G_{\rm RP}$ bands \citep{GaiaCollaboration2023}. As suggested by \citet{Eastman2019}, reasonable systematic error floors were applied: 0.02 magnitudes for Gaia $G$, $G_{\rm BP}$, $G_{\rm RP}$, 2MASS $J$, $H$, and $K$ bands, and 0.03 magnitudes for WISE $W1$, $W2$, and $W3$ bands. Gaussian priors on $\feh$ and parallax were applied, with the former obtained from SWEET-Cat \citep{Santos2013sweetcat, Sousa2021} and the latter drawn from Gaia DR3 \citep{GaiaCollaboration2023}. To be conservative, a minimum error of 0.08 dex was adopted for $\feh$. The V-band extinction and its corresponding uncertainty for each system were derived from the 3D dust map by \citet{Green2019} using the \texttt{dustmaps}\footnote{\url{https://github.com/gregreen/dustmaps}} \citep{Green2018} package. If the 3D map was unavailable, an upper limit on the V-band extinction was adopted from \citet{Schlafly2011}\footnote{\url{https://irsa.ipac.caltech.edu/applications/DUST/}}. The priors used in this work are shown in Table~\ref{tab:prior_list}. 

For systems with nearby stellar companions, a two-component or multi-component SED analysis was performed when band-dependent blending is expected. The broad photometric bands used in SED modeling have varying angular resolutions: $\sim 1\dotarcsec$0 for \gaia\, \citep{GaiaCollaboration2023}, $2\farcs0$ for 2MASS bands \citep{Skrutskie2006}, and $6\farcs1$, $6\farcs4$, and $6\farcs5$ for $W1$, $W2$, and $W3$ bands \citep{Wright2010}, respectively. Consequently, photometric measurements from these surveys can be blended if they cannot fully resolve the targets with nearby stellar companions. Therefore, we performed multi-component SED modeling that met the following criteria: (1) the magnitude contrast between the host and the companion is less than 5, with \gaia\, G-band from \citep{GaiaCollaboration2023} prioritized, followed by the \textit{J}-band, \textit{H}-band, and \textit{K}-band from imaging serveries including \cite{Ngo2015}, \cite{Wollert2015}, \cite{Ngo2016}, and \cite{Bohn2020}; and (2) the stellar companion is unresolved in at least one of the six aforementioned bands. 

\begin{figure*}
    \centering
    \includegraphics[width=1\linewidth]{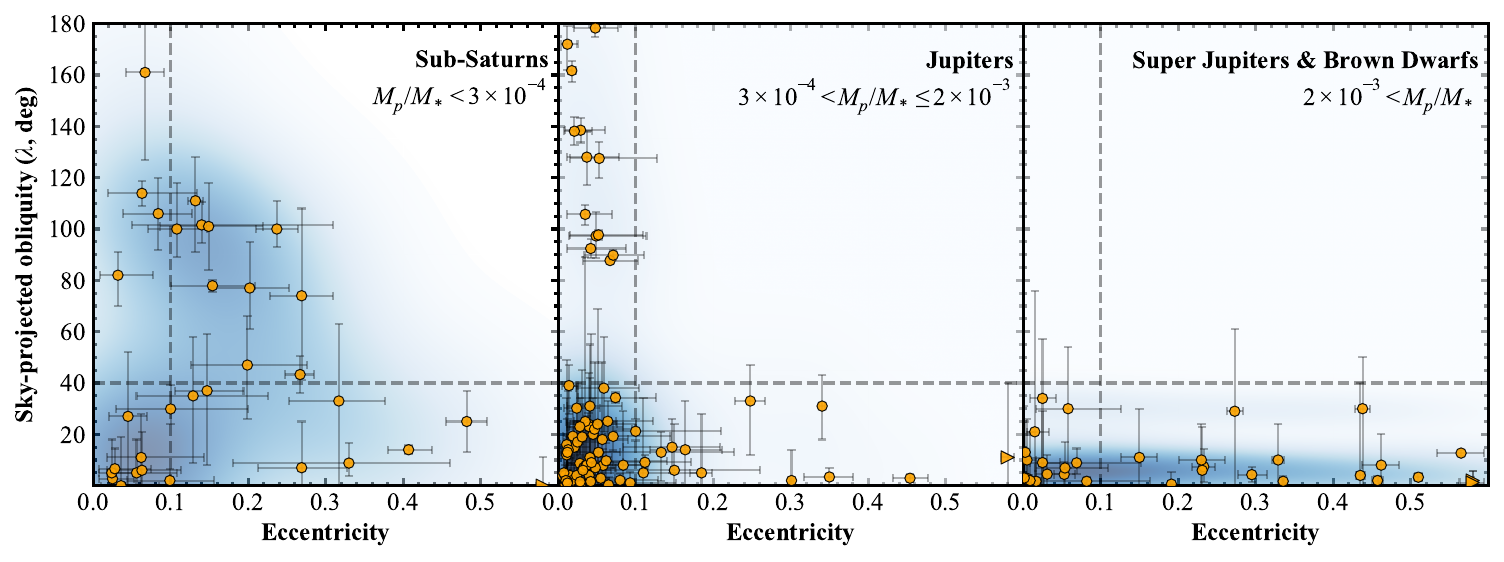}
\caption{
Sky-projected obliquity as a function of orbital eccentricity, shown separately for sub-Saturns (left), Jupiters (center), and more {super Jupiters} and brown dwarfs (right). Orange circles indicate individual systems, while the blue shading represents the underlying density distribution. The vertical dashed lines mark $e = 0.1$, and the horizontal dashed lines represent $\lambda = 40\degree$. Systems with eccentricities $>$ 0.6 are shown at the right boundary with right-pointing triangles.}
\label{fig:elambdarelation}
\end{figure*}

\noindent\textit{\ul{Photometry + Transit Times + RV + RM Modeling}}

We jointly modeled the transit light curves, transit mid-times, radial velocities, and Rossiter-McLaughlin observations. The combined fit provides self-consistent constraints on the transit profiles and ephemerides, orbital RV, as well as  stellar spin and planetary orbit geometry.

For each transiting planet, we adopted uniform priors for the orbital period ($P$), reference transit mid-time ($T_c$), cosine of the orbital inclination ($\cos i$), planet-to-star radius ratio ($\rp / \rstar$), and RV semi-amplitude ($K$). To mitigate the well-known eccentricity bias toward large values \citep{Lucy1971}, we fitted the parameters $\ecosw$ and $\esinw$ instead of eccentricity and argument of periastron separately \citep{Triaud2011, Eastman2013}. When present, additional non-transiting planets in the systems we study are modeled using only the RV-related parameters ($P$, $K$, $\ecosw$, and $\esinw$). Each light-curve dataset had its own fitted flux baseline ($F$) and additional variance term ($\sigma^2$), while each RV dataset had an independent systemic velocity or instrumental offset ($\gamma_{\mathrm{rel}}$), RV variance ($\sigma_J$), and stellar jitter ($\sigma_J^2$). To conservatively account for potential residual contamination in the QLP and SPOC data, we adopted a uniform prior of $\mathcal{U}(0, 1)$ on the dilution factor, while fixing the dilution factor to zero for ground-based transits. If ground-based transits were unavailable, the dilution factor was allowed to vary only for the QLP data. For systems with extremely close stellar companions unresolved by \gaia\ and reported by \cite{Ngo2015,Ngo2016,Bohn2020}, we computed dilution factors from the magnitude differences measured in those imaging surveys. Note that we applied dilution corrections only when the contamination ratio exceeded 1\%.

Four extra parameters were required for the RM fit: the projected stellar rotation velocity ($\vsini$), the projected spin-orbit angle, the microturbulence dispersion ($\xi$), and the macroturbulence dispersion ($\zeta$). Gaussian priors on $\vsini$ were imposed based on literature spectroscopic analyses, with a minimum uncertainty of 0.5 \kms. A uniform prior of $\mathcal{U}(-180, 180)$ was adopted for $\lambda$. Following \cite{Triaud2011,Triaud2018}, we fitted $\sqrt{\vsini}\cos\lambda$ and $\sqrt{\vsini}\sin\lambda$ to avoid biasing $\vsini$ toward higher values. The $\xi$ and $\zeta$ were computed using empirical relations from \cite{Bruntt2010} and \cite{Doyle2014Vmac}, valid within their respective calibration ranges (5000 K $\le$ \teff $\le$ 6500 K, $\log g \ge 4.0$ for \citealt{Bruntt2010}; and 5200 K $\le$ \teff $\le$ 6400 K, $4.0 \le \log g \le 4.6$ dex for \citealt{Doyle2014Vmac}). For stars outside these ranges, we adopted empirical relations calibrated by the Gaia-ESO Survey working groups \citep{Blanco2014HighResolutionSpectralLibrary}. Scaling factors with uniform priors of $\mathcal{U}(0.5, 2)$ for $\xi$ and $\zeta$ were adopted to account for the intrinsic scatter ($\sim1$ \kms) in the empirical calibrations and to mitigate biases in $v\sin i$ that arise when $v\sin i$ is comparable to the macroturbulent velocity. The instrumental broadening $\beta_{\rm IP}$ for each spectrograph was fixed to the value corresponding to its spectral resolution $R$, computed as $\beta_{\rm IP} = c / (2\sqrt{2\ln2} \cdot R)$, where $c$ is the speed of light.

Both ground-based transit photometry and RM observations may require non-constant baselines (e.g., linear, quadratic, or cubic) to account for instrumental systematics and stellar variability. We initially adopt a constant baseline in a joint Transit+RV+RM fit, and subsequently use the Bayesian Information Criterion (BIC, \citealt{Schwarz1978}) to assess whether higher-order baseline models are justified. A more complex baseline model is adopted only if it results in a significant improvement, defined as $\Delta\mathrm{BIC} > 10$ \citep{KassRaftery1995}.

\noindent\textit{\ul{Optimization and Uncertainty estimation}} \\
Optimized initial guesses for fitted parameters can expedite the fitting process. We adopted the Differential Evolution (DE; \citealt{storn1997differential}) global optimization algorithm implemented in \texttt{PyDE}\footnote{\url{https://github.com/hpparvi/PyDE}}, together with the Nelder-Mead simplex local optimization algorithm implemented in \texttt{EXOFASTv2}. The former explores a relatively large parameter space, while the latter refines the parameters based on the results returned by the DE algorithm. 

{To estimate the uncertainties of the system parameters, we used a parallel-tempering Differential Evolution Markov chain Monte Carlo (PT-DE-MCMC, \citealt{braak2006markov}) algorithm, adapted from the IDL implementation in \texttt{EXOFASTv2} and translated into Python. In the fit, the number of temperatures was set to eight, and the number of chains was set to twice the number of fitted parameters. The MCMC procedure was considered converged when the number of independent draws satisfied $T_z > 200$ and the Gelman-Rubin statistic \citep{Gelman1992} was $< 1.1$.}

\noindent\textit{\ul{RM signal-to-noise ratio}} 

{To quantify the detection significance of the RM signal in a uniform way, we compute a matched-filter RM signal-to-noise ratio (SNR) for each RM dataset following \citet{Kipping2024}. The SNR is defined as $\mathrm{SNR}=\sqrt{\Delta\chi^2}$, where $\Delta\chi^2$ is the difference between the best-fit models with and without the RM signal. In this calculation, we use an effective uncertainty of $\sigma_\mathrm{eff}=\sqrt{\sigma^2+\sigma_\mathrm{jit}^2}$. For systems with multiple RM datasets, the individual SNRs are combined in quadrature.
Note that including systems with $\mathrm{SNR}<3$ does not affect our statistical conclusions, although additional high-precision observations would be valuable for further constraining their stellar obliquity measurements.}

In our sample, a small number of systems require special treatment due to data quality or astrophysical complexities (e.g., potentially corrupted SEDs, orbital decay, or grazing transit geometries). These cases are handled individually, with detailed descriptions provided in Appendix~\ref{appendix}. Adopting a misalignment criterion of $\lambda - 3\sigma_{\lambda} > 0$ and $\lambda > 10^\circ$, we find that our results are generally consistent with those reported in the literature and do not alter the classification of most systems.

As an illustrative example, the full modeling results for HD~209458 are shown in Figure~\ref{fig:example_modelling}. {The derived parameters, binary/multiple-star flag, and combined RM SNR for all systems are summarized in Table~\ref{tab:listandparameters}. }

\section{Three populations on the $\MakeLowercase{e - \lambda}$ space}
\label{section:e_lambda}

Our refit analysis provides uniform, self-consistent posteriors and enables a systematic investigation of dependencies among system parameters, particularly the two most informative indicators of dynamical history among the currently best-characterized exoplanetary census, orbital eccentricity ($e$) and stellar obliquity ($\lambda$).

\noindent \ul{Analysis Sample Construction}

Starting from the homogeneous modeling sample described in Section~\ref{sec:Modeling Sample Construction and Data Collection}, we apply the following additional selection criteria to construct the analysis sample used in this section: (1) to avoid planetary-type misclassification, we remove \nbadplmass\, systems with poorly constrained planetary masses, defined by $\mplanet - 3\sigma_{\mplanet} < 0$, (2) Stellar multiplicity has long been theorized as a potential driver of spin-orbit misalignments \citep{Holman1997, Wu2003, Fabrycky2007, Naoz2012}. Recent obliquity studies confirm that warm Jupiters tend to be aligned in single-star systems \citep{Wang2021, Rice2022WJs_Aligned, Wang2024, Espinoza2023} yet are frequently (and sometimes famously, e.g., HD~80606; \citealt{Moutou2009,Pont2009,Winn2009,Hebrard2010}; TIC~241249530; \citealt{Gupta2024}) misaligned in binaries; {super Jupiters} show a similar single-versus-binary contrast \citep{Rusznak2025}. Moreover, hot Jupiters orbiting stars cooler than the Kraft break ($\sim 6500\,{\rm K}$) are rarely misaligned in single-star systems but are frequently misaligned in binary and multi-star systems \citep{Wang2025kb}. Motivated by this, \nbin\, binary and multiple star systems were removed by cross-matching our sample with the Binary and Multiple Star Systems Catalog \citep{Schwarz2016}, the Catalog of Exoplanets in Visual Binaries \citep{Fontanive2021}, the \textit{Gaia} binary catalog \citep{ElBadry2021}, and the NASA Exoplanet Archive \citep{Christiansen2025}. {After applying these selection criteria, the final sample used for the analysis consists of \nsysusedforanalysis\ systems: \nss\ sub-Saturns, \njup\ Jupiters, and \nmassive\ super-Jupiters and brown dwarfs.} Of these systems, \nsysusedforanalysisworm\ (\nssworm\ sub-Saturns, \njworm\ Jupiters, and \nmjworm\ {super Jupiters}) either have obliquities derived from line-by-line RM techniques or lack available RM time-series data. For these systems, we model only the system parameters and do not re-fit the line-distortion data products; therefore, we adopt the literature values of the projected spin-orbit angle. The statistical significance of our results is insensitive to this choice; excluding these systems yields consistent conclusions.

\noindent \ul{Unambiguous three different populations}

A key advance in recent stellar-obliquity studies is the demonstration by \citet{Rusznak2025} that sub-Saturns, Jupiters, and {super} Jupiters/brown dwarfs occupy distinct host-star $\teff - \lambda$ regimes: sub-Saturns can be misaligned around cool stars, hot Jupiters tend to be misaligned only around hot stars, while {super Jupiters} and brown dwarfs remain aligned regardless of the host star's effective temperature (\teff). Using our refined system parameters, we find strong evidence that these populations also exhibit distinct 
 $e - \lambda$  behavior (Figure~\ref{fig:elambdarelation}):

\begin{itemize}
    \item {Sub-Saturns} ($\mratio \leq 3 \times 10^{-4}$, or $M_{\rm p}\leq\sim0.3M_{\rm J}$): Misalignments occur across the full eccentricity range;
    \item {Jupiters} ($3 \times 10^{-4} < \mratio \leq 2 \times 10^{-3}$, or $\sim0.3M_{\rm J}<M_{\rm p}\leq\sim 3M_{\rm J}$): Misalignment confined to circular orbits;
    \item {Super Jupiters and brown dwarfs}  ($\mratio > 2 \times 10^{-3}$, or $M_{\rm p} > \sim3M_{\rm J}$): Aligned for all $e$.
\end{itemize}

\noindent \ul{Statistical Test}

We perform a two-dimensional Kolmogorov--Smirnov (K--S) test \citep{Peacock1983,Fasano1987,press2007numerical},
comparing the joint $(e,\lambda)$ distributions of i) sub-Saturns vs. Jupiters, ii) Jupiters vs. {super Jupiters}/brown dwarfs and, iii) sub-Saturns vs. {super Jupiters}/brown dwarfs, implemented with the public code \texttt{ndtest}\footnote{\url{https://github.com/syrte/ndtest}}. For each pairwise comparison, we propagate measurement uncertainties in $e$ and $\lambda$ with a Monte Carlo resampling scheme to obtain a distribution of $p$-values. We model the (generally asymmetric) uncertainties in $e$ and $\lambda$ for each system with a skewed Gaussian probability density function, providing an analytical representation of the measurement distribution. We then generate $10{,}000$ synthetic realizations by independently drawing $(e,\lambda)$ for every system from these distributions and computing the 2D K--S $p$-value each time. The fractions of realizations with $p<0.05$ are {100}\% for sub-Saturns versus Jupiters, 100\% for Jupiters versus {super Jupiters} and brown dwarfs, and {99}\% for sub-Saturns versus {super Jupiters} and brown dwarfs, suggesting statistically significant differences in the three comparisons. 

{We caution, however, that $\teff$ is not uniformly sampled across the three mass regimes. This is particularly important for sub-Saturns: the $e$--$\lambda$ pattern inferred for this population is based almost exclusively on cool-star systems. This paucity likely reflects both the likely low occurrence rate of sub-Saturns around hot stars \citep{Yang2020} and the difficulty of measuring their stellar obliquity signals \citep[see][for details]{Dugan2025}.
 Therefore, the comparison between misaligned sub-Saturns and misaligned Jupiters, which occur predominantly around hot-star hosts, is not yet a controlled comparison within the same $\teff$ regime. Nevertheless, the difference in $\teff$ alone does not provide an obvious explanation for why misaligned sub-Saturns would retain measurable eccentricities whereas misaligned Jupiters around hot-star hosts would not.}

\section{Discussion}
\label{sec:discussion}

For many years, stellar obliquity has been treated as a master diagnostic for migration histories: measure $\lambda$ well enough and the rest will follow (as reviewed by \citealt{WinnFabrycky2015, Triaud2018, Albrecht2022}). Numerous studies \citep[See, e.g.,][]{FabryckyWinn2009, Knudstrup2024} have drawn inferences along these lines, typically linking large $\lambda$ to violent high-$e$ migration, and more modest, aligned configurations to disk-driven migration. In this work, we revisit that premise. We homogeneously refit an existing catalog of exoplanet systems with Rossiter-McLaughlin effect measurements to obtain robust estimates for the full set of orbital and physical parameters. We then analyzed the resulting $\lambda$ distribution with explicit attention to its dependence on both stellar and planetary properties, in particular planetary eccentricity $e$ and planetary mass $m$. The central outcome is that $\lambda$ depends strongly on \emph{both}. This joint dependence is surprising and places considerable pressure on existing theoretical interpretations that rely on $\lambda$ alone.

The first major tension is with high-$e$ migration scenarios. A broad class of such mechanisms (see \citealt{Dawson2018}, for a review of theories for the high-e origins of hot Jupiters), beginning with the Kozai picture\footnote{In our context, specifically the \emph{planet-planet} variant \citep{Petrovich2016,Naoz2016}, since we restrict attention to systems without confirmed stellar companions.}, predicts substantial spin-orbit misalignments with relatively well-defined characteristic distributions \citep{Beauge2012,Petrovich2016} and has often treated residual eccentricity as a signpost of their operation. Our results point in the opposite direction for Jovian-class and more massive companions: within these regimes, $e$ is \emph{anti-correlated} with $\lambda$. Taken at face value, this pattern is difficult to reconcile with high-$e$ migration as a dominant pathway. 

We further remark that although coplanar high-$e$ migration \citep{Li2014EccentricityGrowth, Petrovich20152015CHEM, Bhaskar2026} successfully predicts the existence of aligned Jupiters on substantially eccentric orbits, {it is, at face value, unclear why this mechanism would seem to operate much more readily in the massive-Jupiter and brown-dwarf regimes than in the Jovian regime.} That said, even if high-$e$ migration is responsible for the high obliquities, additional physics is required to explain why some misaligned sub-Saturns retain non-zero eccentricities, whereas all misaligned hot Jupiters have circularized, despite sub-Saturns being expected to dissipate tidal energy more efficiently \citep{Correia2020, Batygin2025, Gao2025, Petrovich2026}.

A distinct class of proposals invokes primordial excitation of obliquity. Variants include magnetic or gravitational star-disk torques \citep{Foucart2011, Lai2011, Romanova2013, Romanova2021, Batygin2012, Su2025}, chaotic accretion \citep{Bate2010,Thies2011,Fielding2015,Bate2018}, or internal wave-induced stellar tumbling \citep{Rogers2012} that tilt the stellar spin without requiring large $e$. In their simplest forms, these mechanisms operate at the star-disk level and, to leading order, are not expected to depend sensitively on the mass of a single planet or on its orbital state (e.g., orbital eccentricity). Our trends therefore suggest that any successful primordial process must either introduce planet-mass-dependent disk coupling or admit later dynamical modulation that modifies $e$ without substantially reorienting the star. It is tempting to speculate that unmodeled planet-disk interactions could provide such a mass handle; notably, the transitions we identify coincide with familiar changes in disk-planet torque regimes, such as the neighborhood of the thermal mass and, at higher $m$, the approach to deep gap opening. However, quantitative models that connect these regime changes to the observed $\lambda(e,m)$ structure have not yet been developed in sufficient detail for rigorous comparison.

It is also plausible that multiple processes contribute in combination. What our analysis clarifies is that the mixture is \emph{not} uniform across mass, and that $\lambda$ cannot be used in isolation as a reliable tracer of migration history. The relative roles and operating fractions of the underlying mechanisms remain to be established (see \citealt{Esposito2026} for a recent related effort). Cumulatively, the trends reported here underscore the absence of a unified theoretical framework for the origins of spin-orbit misalignment, and they imply that strong inferences about the migratory histories of close-in planets drawn solely from $\lambda$ are premature.

\begin{deluxetable*}{lcccccccc}
\tablecaption{Planetary and Stellar Parameters\label{tab:listandparameters}}
\tablewidth{0pt}
\tablehead{
\colhead{Planet} & \colhead{$\lambda$ (deg)} & \colhead{$T_{\rm eff}$ (K)} & \colhead{$M_*$ ($M_\odot$)} & \colhead{$P$ (days)} & \colhead{$M_p$ ($M_{\rm Jup}$)} & \colhead{$a/R_*$} & \colhead{$e$} & ... \\
}
\startdata
HD 209458 b & $0.49\pm0.30$ & $6107^{+66}_{-62}$ & $1.145^{+0.046}_{-0.050}$ & $3.524749025(15)$ & $0.6913^{+0.0062}_{-0.0070}$ & $8.953^{+0.078}_{-0.061}$ & $0.0089\pm0.0047$ & ... \\
... & ... & ... & ... &... & ... & ... & ... & ... \\
\enddata
\tablenotetext{}{\hspace{.5cm} Note: The full table is \href{https://github.com/wangxianyu7/Data_and_code/tree/main/e-\%CE\%BB_Trends}{available} online. }
\end{deluxetable*}
\section{Acknowledgements}
\label{Acknowledgements}

\begin{deluxetable*}{lccccccc}
\tablecaption{Summary of Data Sets Used in This Work \label{tab:data}}
\tablehead{
\colhead{System} & 
\colhead{Inst} & 
\colhead{Sector} & 
\colhead{Cadence} & 
\colhead{Transit Source} & 
\colhead{Data Type} 
}
\startdata
HD 209458   & TESS      & 56 & 200 s   & TESS-SPOC        & LC \\
HD 209458   & TESS      & 82 & 200 s   & QLP        & LC \\
...  & ...    & ...  & ...  & ...   & ...   \\
\\
\hline
  & Inst      & Telescope   &    N     &     References       &  Data Type  \\
\hline
HD 209458   & CORALIE      & Euler 1.2-m         &  141    & \cite{Naef2004}      & RV \\
HD 209458   & ELODIE       & OHP 1.93-m          &  124    & \cite{Naef2004}      & RV \\
HD 209458   & ESPRESSO     & ESO VLT 8.2-m       &  170    & \cite{Santos2020}    & RV \\
HD 209458   & HARPS        & ESO La Silla 3.6-m  &   29    & DACE                  & RV \\
HD 209458   & HIRES        & Keck I 10-m         &  105    & \cite{Butler2017}    & RV \\
HD 209458   & Hamilton     & Lick 3-m            &   26    & \cite{Rosenthal2021} & RV \\
...         & ...          & ...                 & ...     & ...                  & ... \\
\enddata
\tablecomments{Number of RV points and TESS/Kepler sectors are shown. Cadence corresponds to the light curve time sampling. DACE stands for Data and Analysis Center for Exoplanets (\url{https://dace.unige.ch}). The full table is \href{https://github.com/wangxianyu7/Data_and_code/tree/main/e-\%CE\%BB_Trends}{available} online.{ The references listed in this table have been included in the bibliography through \texttt{\textbackslash nocite} to provide proper credit to the original studies.} }
\end{deluxetable*}
\nocite{2001ApJ...552..699B,2004ApJ...613L.153A,2005ApJ...633..465S,2006ApJ...636..445C,2006ApJ...650.1160B,2006ApJ...651L..61O,2007A&A...472L..13G,2007AJ....134.1707W,2007ApJ...657.1098W,2007ApJ...658.1322C,2007ApJ...666L.121T,2007ApJ...670L..41K,2007ApJ...671.2115B,2007MNRAS.375..951C,2007PASJ...59..763N,2008A&A...482L..21A,2008A&A...491..889D,2008ApJ...675.1531W,2008ApJ...675L.113W,2008ApJ...677..657J,2008ApJ...682.1283W,2008ApJ...683.1076W,2008ApJ...686..649J,2008ApJ...689L.149C,2009A&A...496..259G,2009A&A...501..785G,2009A&A...502..391S,2009A&A...502..395W,2009A&A...503..601B,2009A&A...506..377T,2009AJ....137.4911F,2009ApJ...690L..89H,2009ApJ...691.1145S,2009ApJ...693.1920H,2009ApJ...694.1559S,2009ApJ...700..302W,2009ApJ...703L..99W,2009ApJ...704.1107L,2009ApJ...706..785H,2009ApJ...707..167S,2009MNRAS.392.1532J,2009MNRAS.396.1023S,2009MNRAS.399..287S,2009PASJ...61..991N,2009PASP..121.1104J,2010A&A...511A...3G,2010A&A...519A..98B,2010A&A...520A..56S,2010A&A...524A..25T,2010A&A...524A..55G,2010AJ....139..176F,2010AJ....140.2007M,2010ApJ...709..159A,2010ApJ...710.1724B,2010ApJ...715..421T,2010ApJ...718..575W,2010ApJ...720..337S,2010ApJ...720.1118B,2010ApJ...724..866K,2010MNRAS.405.1867S,2010MNRAS.407..507C,2010MNRAS.407.2625M,2010MNRAS.408.1680S,2010PASJ...62..653N,2010PASJ...62L..61N,2011A&A...525A..54B,2011A&A...527A...8S,2011A&A...527A..85N,2011A&A...527L..11H,2011A&A...531A..40F,2011A&A...531A..60A,2011A&A...533A..88G,2011A&A...533A.113M,2011A&A...534A..16A,2011A&A...535L...7H,2011AJ....141...63W,2011AJ....142...86E,2011ApJ...726...52H,2011ApJ...726L..19A,2011ApJ...730L..31H,2011ApJ...733..116B,2011ApJ...733..127S,2011ApJ...735...24J,2011ApJ...738...50A,2011ApJ...742...59H,2011ApJ...742..116B,2011MNRAS.414.3023S,2011MNRAS.417..709A,2011PASJ...63L..57H,2011PASJ...63L..67N,2011PASJ...63S.531H,2012A&A...546A..27B,2012A&A...547A..61S,2012AJ....144...19B,2012AJ....144..139H,2012ApJ...744..189A,2012ApJ...746..111T,2012ApJ...750...84B,Albrecht2012,2012ApJ...759L..36H,2012ApJ...760..139B,2012ApJ...761..123S,2012MNRAS.420.2580S,2012MNRAS.422.1988A,2012MNRAS.423.1503B,2012MNRAS.426..739H,2012MNRAS.426.1338S,2013A&A...549A..18T,2013A&A...549A.134H,2013A&A...551A..11M,2013A&A...551A..73F,2013A&A...551A..80T,2013A&A...552A...2L,2013A&A...552A..82G,2013A&A...552A.120S,2013A&A...559A..32N,2013AJ....146..113B,2013AJ....146..147M,2013ApJ...770...95F,2013ApJ...773...64P,2013ApJ...774L...9A,2013ApJ...775...54S,2013MNRAS.428.3671T,2013MNRAS.430.2932M,2013MNRAS.434.1300S,2013PASP..125...48M,2014A&A...562A.126M,2014A&A...562L...3G,2014A&A...564L..13E,2014A&A...568A..81L,2014A&A...570A..64S,2014A&A...572A..49N,2014AJ....147...39C,2014AJ....147...84B,2014AcA....64...27M,2014ApJ...785..126K,2014MNRAS.440.1982H,2014MNRAS.440.3392B,2014MNRAS.443.2391M,2014MNRAS.444..776S,2014MNRAS.445.1114A,2014arXiv1410.3449A,2014arXiv1412.7761B,2015A&A...575A..61A,2015A&A...575L..15S,2015A&A...577A..54C,2015A&A...577A.109M,2015A&A...579A..55B,2015A&A...579A.136M,2015A&A...580A..63M,2015AJ....150...12B,2015AJ....150...18H,2015AJ....150...85H,2015AJ....150..168H,2015ApJ...800L...9A,2015ApJ...812L..11S,2015ApJ...813L...9D,2015ApJ...814L..16Z,2015MNRAS.447..711S,2015MNRAS.450.1760T,2015MNRAS.450.2279T,2015PASP..127..143R,2016A&A...585A.126W,2016A&A...586A..93N,2016A&A...591A..55M,2016A&A...594A..65N,2016AJ....151...17J,2016AJ....151...45E,2016AJ....151..171J,Zhou2016,2016ApJ...818...46M,2016ApJ...820...87V,2016ApJ...823...29A,2016ApJ...824...55S,2016ApJ...825...53H,2016MNRAS.457.4205S,2016MNRAS.458.4025D,2016MNRAS.459.4281K,2016PASP..128f4401T,2016PASP..128l4402B,2017A&A...599A...3L,2017A&A...600L..11C,2017A&A...601A..53E,Bonomo2017,2017A&A...604A.110A,2017A&A...606A..73T,2017AJ....153...94C,2017AJ....153..131N,2017AJ....153..200A,Butler2017,2017AJ....153..211Z,2017AJ....153..215P,2017AJ....153..263M,2017AJ....154...49W,2017AJ....154..107P,2017AJ....154..122C,2017AJ....154..237V,2017AN....338...35G,Brown2017,2017MNRAS.465..843M,2017MNRAS.465.3693H,2017MNRAS.467.1714T,2017PASJ...69...29N,2018A&A...610A..63D,2018A&A...613A..41M,2018A&A...619A...1B,2018AJ....155...52A,Brady2018,2018AJ....156..181W,2018AJ....156..197A,2018AJ....156..250Y,Maciejewski2018AcA68371M,2018IBVS.6243....1M,2018MNRAS.475.1765B,2018MNRAS.475.1809G,2018MNRAS.477.2572B,2018MNRAS.478.4866V,2018MNRAS.478.5356S,2018MNRAS.480.5307T,2018MNRAS.481..596J,2018MNRAS.481.4960R,2018arXiv180907709A,2018arXiv181209264A,2019A&A...622A..81M,2019A&A...630A..81B,2019AJ....157...31Z,2019AJ....157...74A,2019AJ....157...82W,2019AJ....157..100J,2019AJ....157..141T,2019AJ....158...78J,2019AJ....158..141Z,2019AJ....158..197R,2019MNRAS.482.1379H,2019MNRAS.482.1807K,2019MNRAS.484.3522H,2019MNRAS.485.5168M,2019MNRAS.488.3067H,2019MNRAS.490.1479H,2019MNRAS.490.2467T,Mancini2020,2020A&A...635A..60D,2020A&A...635A.205B,2020A&A...636A..98C,2020A&A...642A..50L,2020AJ....159...44C,2020AJ....159..145J,2020AJ....159..243P,2020AJ....160..114C,2020AJ....160..179M,2020AJ....160..192S,2020AJ....160..222J,2020AJ....160..235B,2020MNRAS.494..750E,2020MNRAS.494.5872B,2020MNRAS.499..428S,2020Natur.580..597E,2021A&A...645A..71C,2021A&A...646A.168C,2021A&A...650A..66B,2021A&A...653A.104B,2021A&A...654A..73S,Sousa2021,2021AJ....161...82S,2021AJ....161..119R,2021AJ....161..194R,2021AJ....162...54H,2021AJ....162..182R,2021AJ....162..256W,2021AJ....162..292A,2021ApJ...917L..34W,Dong2021,2021ApJS..254...39G,Rosenthal2021,2021ApJS..255...15W,2021MNRAS.505.4956B,2021MNRAS.507.4132M,2021PNAS..11817418H,2022A&A...662A.107S,2022A&A...663A.160B,2022A&A...664A.162M,2022A&A...666A..47Z,2022A&A...667A...1S,2022A&A...667A..22K,2022A&A...667A.127M,2022A&A...668A..31B,2022AJ....163....9I,2022AJ....163...82W,2022AJ....163...89C,2022AJ....163..133E,2022AJ....163..158W,2022AJ....163..197B,2022AJ....163..225T,2022AJ....163..227A,2022AJ....164...50C,Rice2022WJs_Aligned,2022AJ....164..178P,2022AJ....164..198M,2022ApJ...926L...7D,2022ApJ...931L..15S,2022MNRAS.513.5955K,2022MNRAS.514.4944C,2022MNRAS.515.1247T,2022MNRAS.516..636S,2023A&A...669A..40O,2023A&A...671A.163M,2023A&A...671A.164K,2023A&A...672A.134S,2023A&A...673A..42N,2023A&A...675A..39P,2023A&A...679A..70H,2023AJ....165...60V,2023AJ....165...65R,2023AJ....165..207C,2023AJ....165..234G,2023AJ....165..268V,2023AJ....166...30C,2023AJ....166..136D,2023AJ....166..159H,2023AJ....166..163H,2023AJ....166..217W,2023AJ....166..225S,2023AJ....166..271E,2023AcA....73...57M,2023ApJ...944L..41F,2023ApJ...949L..35H,2023ApJ...951L..29D,Espinoza2023,2023ApJ...959L...5L,2023ApJS..265....1Y,2023MNRAS.521.2765R,2023NatAs...7..198Z,2023Natur.623..932L,2024A&A...682A.135C,2024A&A...684L..17M,2024A&A...685A..90M,2024A&A...686A.147Z,Sicilia2024,2024A&A...687L...2Z,2024A&A...690A..18C,Knudstrup2024,2024A&A...691A.120S,2024AJ....167..175H,2024AJ....168...81S,2024AJ....168..116R,2024AJ....168..145F,2024AJ....168..185E,2024AJ....168..194T,2024AJ....168..295Z,2024ApJ...962L..22D,Wang2024,2024ApJS..270....8W,2024ApJS..272...32P,2024MNRAS.52710955B,2024MNRAS.532.1612H,2024NatAs...8..909B,Gupta2024,2025A&A...694A..36H,2025A&A...694A..91Z,TalaPinto2025,2025A&A...700A...5G,2025A&A...702A.266Z,2025AJ....169....4D,2025AJ....169..212H,2025AJ....169..225Y,2025AJ....170...34T,2025AJ....170...51T,2025AJ....170...70E,2025AJ....170..175Z,2025AJ....170..274Z,2025AJ....170..313W,Dugan2025,2025ApJS..280...30Y,2025MNRAS.536.3745D,2025PASP..137g4401G,2026AJ....171...12C,2026ApJ...996L..13E,Gan2024}
\nocite{2017A&A...601A.117G}
\nocite{2022A&A...658A..75H}
\nocite{2026ApJ..1000L..54Y,2026A&A...708A.272V,2026AJ....171..139B,2025A&A...704A.194M,2025AJ....170..275Y,2025AJ....170..182P,2025MNRAS.536.2581J,2024AJ....168..196L,2024ApJ...971L..40R,2024A&A...688A.172M,2024A&A...686A.301K,2024A&A...682A.129M,2024A&A...682A..28C,2023ApJ...959L...1H,2023AJ....166..130S,2023A&A...677A.110J,2022MNRAS.516..298D,2022AJ....164...21S,2022A&A...660A..52C,2022AJ....163..101L,2021MNRAS.505..435S,2021AJ....161...70P,2021AJ....161...18M,2020A&A...642A..54C,2020A&A...635A.171C,2019A&A...631A..76H,2019A&A...631A..28D,2017A&A...608A.135C,2016AJ....151..138R,2015AJ....150..197H,2015A&A...581L...6D,2015A&A...578L...4L,2015A&A...575A.111D,2014A&A...571A..37S,2014ApJ...792..112A,2013ApJ...772...80F,2013ApJ...771...11A,2012ApJ...749..134H,2012A&A...537A.136G,2011A&A...533A.130H,2011PASP..123..547M,2010ApJ...725.2017K,2010ApJ...724.1108J,2010ApJ...723L.223W,2010ApJ...722L.224B,2010A&A...517L...1Q,2009PASJ...61L..35N,2009ApJ...696.1950B,2009ApJ...693..794W,2009ApJ...690.1393S,2008A&A...488..763H,2008ApJ...683L..59C,2008MNRAS.387L...4A,2008A&A...482L..25B,2008A&A...482L..17B,2008PASJ...60L...1N,2008A&A...481..529L,2008ApJ...673L..79N,2007ApJ...667L.195M,2007ApJ...667..549W,2007ApJ...656..552B,2006ApJ...653L..69W}
 
\begin{deluxetable*}{cccccccc}
\tablecaption{Priors \label{tab:prior_list}}
\tabletypesize{\scriptsize}
\tablehead{
\colhead{System} & \colhead{\feh} & \colhead{Ref.} & \colhead{\vsini} & \colhead{Ref.} & \colhead{Parallax} & \colhead{Max A$_{v}$}\\
\colhead{} & \colhead{(dex)} & \colhead{ } & \colhead{\kms} & \colhead{} & \colhead{mas}
}
\startdata
HD 209458 & $\mathcal{N}(0.04, 0.01)$  & \cite{Sousa2021} & $\mathcal{N}(4.49,0.5)$ & \cite{Bonomo2017} & $\mathcal{N}(20.80312, 0.02842)$ & 0.0  \\
... & ... & ... & ... & ... & ... &...\\
\enddata
\tablecomments{This table is only partially displayed here to illustrate its format and content. The complete, machine-readable version of the table is available in the electronic version of this work. The full table is \href{https://github.com/wangxianyu7/Data_and_code/tree/main/e-\%CE\%BB_Trends}{available} online. { The references listed in this table have been included in the bibliography through \texttt{\textbackslash nocite} to provide proper credit to the original studies.}}
\end{deluxetable*}

{We sincerely thank the reviewer for their constructive comments, which have helped us improve the manuscript.}
We are grateful for helpful discussions with Jason Eastman, Cristobal Petrovich, Malena Rice, Gongjie Li, Yubo Su, Yanqin Wu, Jack Lubin, Janosz Dewberry, James Owen, Simon H. Albrecht, and, Joshua N. Winn. This work was supported in part by the NASA Exoplanets Research Program  NNH23ZDA001N-XRP (Grant No. 80NSSC24K0153), the NASA TESS General Investigator Program, Cycle~7, NNNH23ZDA001N-TESS (Grant No. 80NSSC25K7912), and the Heising-Simons Foundation (Grant \#2023-4050). We acknowledge funding support from grant JWST-GO-09025.010-A provided by NASA via the Space Telescope Science Institute under the JWST General Observers Program \#9025. The Space Telescope Science Institute is operated by the Association of Universities for Research in Astronomy, Inc., under NASA contract NAS 5-03127 for JWST. X.Y.W. acknowledges support from the Sullivan Prize Fellowship. S.W. gratefully acknowledges support from the John and A-Lan Reynolds Faculty Research Fund, which enabled participation in the International Conference on Exoplanets and Planet Formation (EPF). This support was vital for discussing the result with the world's leading researchers in the field. Additionally, this research was supported in part by Lilly Endowment, Inc., through its support for the Indiana University Pervasive Technology Institute.

{All the \kepler\, \citep{KeplerLCSC}, \ktwo\, \citep{Vanderburg2014K2SFF,10.17909/t9bc75}, \tess\, QLP \citep{QLP}, and \tess\, SPOC \citep{SPOC} data used in this paper can be found in the Barbara A. Mikulski Archive for Space Telescopes (MAST) \footnote{\url{https://archive.stsci.edu}}.}

\facilities{\kepler, {\it Spitzer}, \tess}

  \software{
  \texttt{allesfast} \citep{allesfitter-paper,allesfitter-code},
  \texttt{Allesfitter} \citep{allesfitter-paper,allesfitter-code},
  \texttt{barycorrpy} \citep{barycorrpy},
  \texttt{dustmaps} \citep{Green2018},
  \texttt{emcee} \citep{emcee},
  \texttt{EXOFASTv2} \citep{Eastman2017, Eastman2019},
  \texttt{lightkurve} \citep{Lightkurve2018},
  \texttt{matplotlib} \citep{hunter2007matplotlib},
  \texttt{ndtest} \citep{Li2025_ndtest},
  \texttt{numpy} \citep{oliphant2006guide, walt2011numpy, harris2020array},
  \texttt{pandas} \citep{mckinney2010data},
  \texttt{PyDE}\footnote{\url{https://github.com/hpparvi/PyDE}},
  \texttt{scipy} \citep{virtanen2020scipy},
  \texttt{wotan} \citep{Hippke2019}
  }

\appendix
\section{Special cases and modeling considerations}
\label{appendix}

\noindent\uline{{Bimodality in stellar mass and age posteriors}}

{SED+MIST modeling can yield bimodal posteriors in stellar mass and age when a star lies near an evolutionary transition point, such as the main-sequence turnoff or the subgiant branch. Similar bimodal solutions have been reported for TOI-569 \citep{Carmichael2020}, TOI-3894 and TOI-5301 \citep{Schulte2024meepI}, and TOI-4138 \citep{Schulte2025meepII}. In our sample, 14 systems exhibit such bimodal behavior: HAT-P-33, HAT-P-41, HAT-P-50, HAT-P-67, K2-234, K2-261, KELT-6, KELT-11, TOI-1842, TOI-2025, TOI-2145, TrES-4, WASP-13, and WASP-38.
}
{We identify bimodality using the Gaussian kernel-density estimate of the derived $M_\star$ posterior. Specifically, we locate the two highest local maxima in the KDE and use the intervening local minimum as the boundary between the two modes. A system is flagged as truly bimodal if (1) the lower-mass mode contains at least 10\% of the total posterior probability and (2) the two modes are separated by at least $2\sigma$. Both the lower- and higher-mass solutions are listed in Table~\ref{tab:listandparameters}. For the statistical analysis, we adopt the mode with the higher posterior probability.}

\noindent\uline{Imperfections in modeling fast rotators:}\\
\noindent\uline{HAT-P-2 and HAT-P-67}

For HAT-P-2 and HAT-P-67, the \cite{Hirano2011} RM model does not capture all features of the observed RM signal, leaving significant structured residuals. Both systems host bright, rapidly rotating stars (HAT-P-2: $\vsini = 22,\kms$, $G = 8.6$; HAT-P-67: $\vsini = 36\kms$, $G = 10$). Similar residual structure was reported for HAT-P-2 by \cite{Albrecht2012}, who speculated that it may arise from variations in limb darkening with the depth of stellar absorption lines. This mechanism may also explain the structured residuals observed for HAT-P-67. Notably, applying the empirical RM model to HAT-P-2 \citep{Winn2007}, and the numerical RM model to HAT-P-67 \citep{Sicilia2024}, does not produce such residuals, indicating that additional physical effects may need to be incorporated into the analytical \cite{Hirano2011} RM model for rapidly rotating stars. Nevertheless, the derived $\lambda$ values are consistent with those reported by \cite{Albrecht2012} and \cite{Sicilia2024}.

\noindent\uline{Infrared excess: HAT-P-1}

The exoplanetary system HAT-P-1 has a stellar companion ($G=9.6$) at a separation of $11\farcs3$, which should not contaminate the WISE photometry given the WISE angular resolution of $\sim 6\farcs5$. Nevertheless, in our SED fits, the WISE $W1$, $W2$, and $W3$ fluxes are consistently higher than the best-fit model. We therefore inflated the uncertainties of these bands to 0.1 mag to absorb any residual contamination from the nearby companion and other possible systematics.

\noindent\uline{Possible corrupted SED: KELT-24}

The recent analysis by \cite{Giovinazzi2024} shows that the SED of KELT-24 may be corrupted. Therefore, following \cite{Giovinazzi2024}, we did not include the SED in our global modeling and instead imposed a spectroscopic prior on \teff\ of $\mathcal{N}(6499, 98)$ K. 

\noindent\uline{Degeneracy for high-grazing systems:\\ WASP-174 and TOI-5027}

For systems exhibiting high impact parameters, the transit geometry becomes intrinsically degenerate, making it difficult to precisely constrain the orbital inclination (or equivalently the impact parameter) and $R_{p}/R_{*}$. In our sample, both WASP-174 and TOI-5027 have values of $b$ close to unity.

To regularize the solution and avoid poorly constrained grazing configurations, we impose additional Gaussian priors on $R_{p}/R_{*}$, adopting $\mathcal{N}(0.1098, 0.0030)$ for WASP-174 \citep{Mancini2020} and $\mathcal{N}(0.106, 0.005)$ for TOI-5027 \citep{TalaPinto2025}.

\noindent\uline{Orbital decay: WASP-12}

Due to the significant orbital decay \citep{Patra2017}, the ephemeris for WASP-12 derived from \tess\ data is not suitable for predicting the RM mid-transit time. Therefore,  we instead used ground-based transit observations from \cite{Maciejewski2018AcA68371M} obtained both before and after the RM measurement \citep{Albrecht2012}.

\end{CJK*}
\bibliography{new.ms.bib}
\bibliographystyle{aasjournal}

\end{document}